\documentclass[twocolumn,widetext,amsmath,amssymb]{revtex4}
\usepackage{graphicx}
\usepackage{dcolumn}
\usepackage{bm}
\begin{document}
\title{Exploring the Evolution of London's Street Network in the Information Space: a Dual Approach}
\author{A. Paolo Masucci$^a$}
\author{Kiril Stanilov$^{b,a}$}
\author{Michael Batty$^a$}
\affiliation{%
a-Centre for Advanced Spatial Analysis, University College of London, London, UK;\\
b-Martin Centre, University of Cambridge, Cambridge, UK}
\begin{abstract}
 We study the growth of London's street-network in its dual representation, as the city has evolved over the last 224 years. The dual representation of a planar graph is a content-based network, where each node is a set of edges of the planar graph, and represents a transportation unit in the so-called information space, i.e. the space where information is handled in order to navigate through the city. First, we discuss a novel hybrid technique to extract dual graphs from planar graphs, called the \textit{hierarchical intersection continuity negotiation principle}. Then we show that the growth of the network can be analytically described by logistic laws and that the topological properties of the network are governed by robust lognormal distributions characterising the network's connectivity and small-world properties that are consistent over time. Moreover, we find that the double-Pareto-like distributions for the connectivity  emerge for major roads and can be modelled via a stochastic content-based network model using simple space filling principles. 

\end{abstract}

\date{\today}

\maketitle

\section{Introduction}

Understanding constitutional principles and structural morphology of complex transportation systems is an outstanding problem in statistical physics and complex systems. 
Energy flows in different systems in  different articulate ways. 
Examples are the circulation of blood in vein vessels or the flow of macromolecules transported between cellular components, river networks and fracture patterns \cite{klein,mol,bohn2}. 
Beside the apparent diversity of the aforementioned phenomena, striking regularities emerge between them. 
Depending on the dimensionality of the space where they are embedded, we can observe similar patterns and scaling laws  \cite{bohn2007,banavar} (Fig.\ref{f0}).

 Here we consider urban street networks in a similar way. 
The recognition of network	 complexity as a key theme in urban studies can be traced back to Euler's first graph theory approach \cite{euler}. 
Similar conceptualizations dominated the scientific discourse during the twentieth century, epitomized by Zipf's and Gibrat's laws.
 While the application of these laws to city size and urban growth still remains an open problem \cite{rozenzipf,rozengib},  during the last decades fractal theory and diffusion  processes found fertile applications to morphological studies of urban systems \cite{mikefract}.

The most direct way to represent street networks is via  \textit{planar graphs}, which are defined as a set of vertices and edges $\left\{V,E\right\}$ embedded in a two-dimensional surface, with the condition that the links do not cross one other, the latter known as the \textit{planarity criteria}. 
The description of a street network via its planar graph representation is known as the \textit{primal graph}.
However, it has been shown that such a representation is not sufficient to describe the complexity of street networks. 
Even if primal graphs have a rich geometrical texture, their topological properties are very similar to the ones of random geometric graphs and do not tell us much about the structure and complexity of urban systems.

\texttt{\begin{figure}[!ht]\center
                \includegraphics[width=0.47\textwidth]{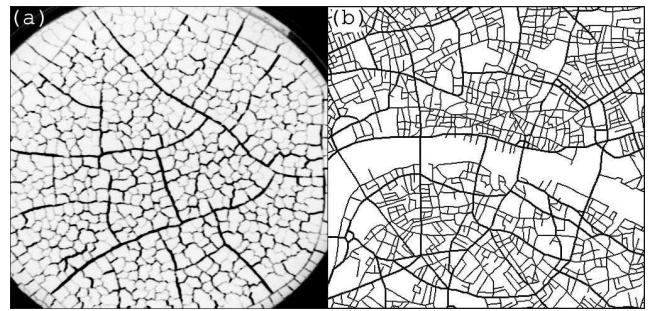}
 \caption{\label{f0} (a) Crack pattern formation from drying a solution of corn-starch and distilled water \cite{mud}. (b) Street network pattern for part of central London around the Thames, where the thick lines represent mayor roads. }
 \end{figure}} 
 
It is now well accepted that such the complexity of urban transport networks resides in the so-called \textit{information space}, or the \textit{dual representation} of the network \cite{rosvaldual}.
Such a representation emerges from the evidence that the primal graph units, i.e. the street segments, are not the constitutive transportation units of urban networks. 
Generally, transportation entities in street networks are constituted not by individual street segments, but by assemblages of such segments, i.e. the roads.
In the field of complex systems, the dual representation of a planar graph is a network where the nodes or vertices are the transportation units, i.e. a collection of street segments belonging to the same road. Then in the dual graph two vertices are linked if the corresponding transportation units intersect  \cite{portadual,jiang}. 

As we have already noted, the dual representation of a street network  can be  interpreted as the \textit{information space} of that network \cite{rosvaldual}. 
This is because the dual representation reflects a system of coding of symbolic information which helps individuals to navigate through urban space. 
Thus a street network can be described by the interplay between two layers, one embedded in the Euclidean space (the primal representation), the other one embedded in a symbolic space (the dual representation).
This is a fertile perspective on spatial networks, since it has been shown that such an interplay can be described as an optimization process that tends to minimize the walker effort both in the Euclidean and in the information space  \cite{masuccilond}.
Moreover striking regularities have been found in the information space of street networks, such as broad connectivity distributions and small-world properties \cite{rosvaldual, kalapala, masuccilond}. 

However, while the primal graph is a straightforward representation, extracting the dual representation of a planar graph is not a trivial problem. The key issue is to determine which street segments belong to the same transportation unit.
 Two main approaches to solve this problem have been proposed reflecting a distinction between physical and behavioural considerations. 
 There are many possible variants of these two approaches, but here we illustrate the first by the \textit{intersection continuity negotiation} (ICN hereafter)  \cite{portadual}, which is based on the geometrical properties of the planar graph. 
 The second method is called the  \textit{street name approach} (SN hereafter) \cite{jiangdual}, and it is based on the symbolic properties of the streets, derived from historical naming conventions. 

 The ICN method, based on the principle that two spatially aligned street segments are likely to belong to the same road, works very well on grid-like street networks, which typically reflect direct top-down planning interventions.
  The ICN method, however, is often misleading when applied to cities  exhibiting a more complex geometry, which have evolved as a result of  bottom-up actions, i.e.  \textit{self-organising cities}. 
A good example illustrating the problems of applying the ICN approach to non-linear geometrical entities is London's orbital motorway M25.
 In the dual representation, such beltways, which are common in many large cities around the world, should be represented by a single vertex, denoting one of the largest hubs in the information space. 
 However,  the ICN method would break the M25, with its circular shape,  into many different vertices,  recognizing as a series of single roads just the street segments that are most aligned. 
On the other hand, the SN method is based on the simple principle that two contiguous street segments that have the same street name belong to the same road. The method is based on the assumption that street naming systems encapsulate the perceptions of how  streets are identified and  used and identified as the main constituent blocks of the city.   SN works very well for large roads, but often, especially in large cities such as London or Paris, the street name could change several times along the same road and as a result the dual representation of the  network based on SN can be quite misleading.
 To overcome the methodological problems of ICN and SN explained above, we introduce a hybrid methodology for extracting the dual graph by mixing the symbolic approach with the geometrical one. 
 
In this paper we present the results of the dual analysis on a unique dataset consisting of nine map series  that record the evolution of Greater London's street network from 1786 to 2010. 
 The dataset is shown in Fig.\ref{f1} and the extraction procedures and its primary representation analysis are described in  \cite{masucci2}.
Each road in the dataset is classified according to a four-level hierarchy based on motorways, class A and class B roads, and minor roads \footnote{This classification was first introduced in the UK during the 1930s. For the earlier maps, which showed only a generic distinction between major and minor roads, we assigned classes A and B to the major roads in accord with their subsequent classification, but only if the road is connected to another major road.}. 
 
We find that the growth of the city in the information space follows rules that are similar to the ones in the euclidean space, i.e. the growth of the network in terms of vertex and edge dynamics can be analytically calculated via two simple logistic laws. Moreover, we conclude that the scale-free connectivity distribution described in other studies is an effect of the ICN method  and does not reflect the true nature of the information space \cite{rosvaldual,portadual,masuccilond}. 
In particular, we show that with an appropriate definition of the dual graph the connectivity distribution comes out to be a robust log-normal distribution over more than two orders of magnitude. However, we find that the small-world properties of the network is a stable attribute of London's street network throughout the last two centuries, highlighting that the navigability of the city in the information space is a robust property of the system.

 Furthermore, we find that the connectivity for the dual network formed just by major roads follows a double-Pareto distribution. 
This property, which was already observed in a study of large-scale national road networks in three different countries \cite{kalapala},  seems to be a peculiar feature of street networks formed by high hierarchy roads as well.  
We show that such a behaviour can be reproduced by a space filling model, where the street segments are the results of a fragmentation process of longer street segments, while longer roads can be composed of street segments created at different times.


\texttt{\begin{figure}[!ht]\center
                \includegraphics[width=0.5\textwidth]{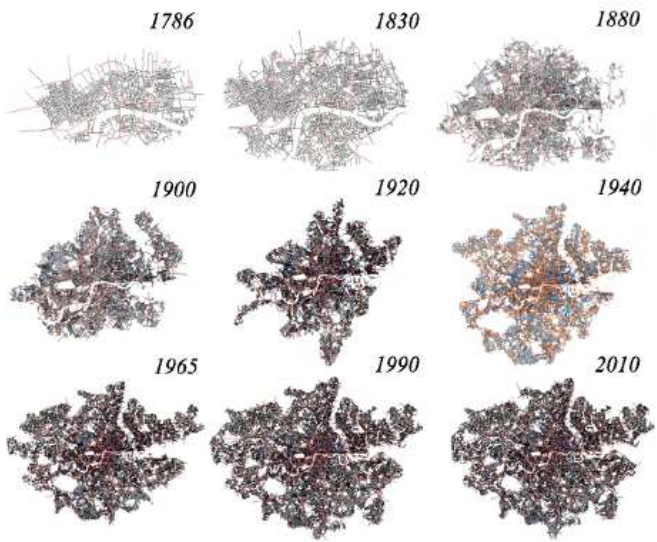}
 \caption{\label{f1} (Colours on-line) The street network of London from 1786 to 2010. Different colors correspond to different road classifications: motorways, A (red) and B (blue) roads, minor roads (grey) (image from \cite{masucci2}).}
 \end{figure}} 

\section{Hierarchical Intersection Continuity Negotiation}

Each street segment in our map series is classified according to four hierarchical levels which broadly reflect capacity: motorways, class A roads, class B roads, and minor roads (see Fig.\ref{f1}). As shown in \cite{masucci2}, the major roads (the motorways, class A and class B roads) and the minor roads reflect two different aspects of the street network. The major roads are conduits for the main inter and intra-city flows of people and resources, while the minor roads serve to access and develop specific plots  of land for various  types of urban (primarily residential) uses. 
 
 In order to extract a reliable dual graph, we combine the geometrical ICN  approach with a symbolic approach, that uses the hierarchical tags of the street segments. We apply  the ICN principle \footnote{``All the nodes are examined in turn. At each node, the continuity of street identity is negotiated among all
pairs of incident edges: the two edges forming the largest convex angle are assigned the highest continuity
and are coupled together; the two edges with the second largest convex angle are assigned the second largest
continuity and are coupled together, and so forth; in nodes with an odd number of edges, the remaining
edge is given the lowest continuity value." from \cite{portadual}}, with a $\pi/2$ threshold (two street segments forming an angle less than $\pi/2$ cannot belong to the same road, see Conclusions for further discussion about this threshold)   first to motorways, then to A roads, and then to B roads. 
 After that, we apply the ICN principle to minor roads, with the rule that two street segments with  the minor road tag cannot belong to the same transportation unit if they are on the opposite sides of a major road.
 
Our method of extracting the dual graph, which we call \textit{hierarchical intersection continuity negotiation} (HICN hereafter), is inspired by the study in \cite{pernadual}. 
Its advantage is that it addresses known deficiencies of both the ICN and the SN approaches. 
We found that the use of road hierarchy as a symbolic layer is less restrictive than the street names based approach SN, as it is not affected by arbitrary  street name changes. 
Compared to the classical ICN, our approach avoids the problems associated with long irregular roads as the application of the road hierarchy automatically prevents the merging of minor and major roads. 

\begin{figure}[!ht]\center
                \includegraphics[width=0.5\textwidth]{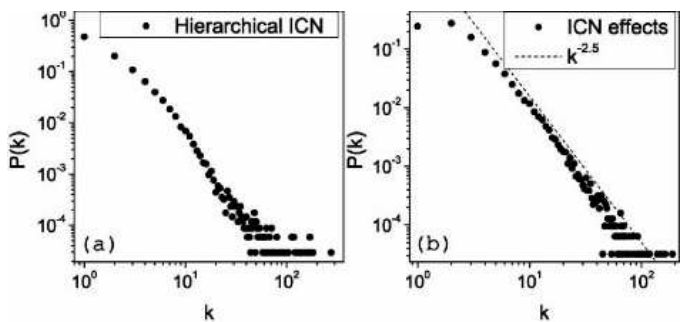}
 \caption{\label{f3} Degree distribution measured for  London's 1990 dual street network: (a) for the network extracted via HICN; (b) for the network extracted via ICN.}
 \end{figure}

 In order to understand to what extent  different dual approaches could affect topological results, we tested the two methods to derive  one of the main topological properties for a complex network, the degree distribution. 
 In Fig.\ref{f3}, we show the degree distribution measured for the dual representation of the 1990 London street network, on a log-log scale. 
The left panel shows  the results  derived by the HICN methodology, while in the right panel  we show the same measure on the dual graph extracted by the ICN principle, with $\pi/2$ angular threshold. 
The systematic errors of ICN, outlined in the introduction, appear to generate a scale-free distribution, while the plot obtained from the HICN does not conform to a power law.

\begin{figure}[!ht]\center
                \includegraphics[width=0.5\textwidth]{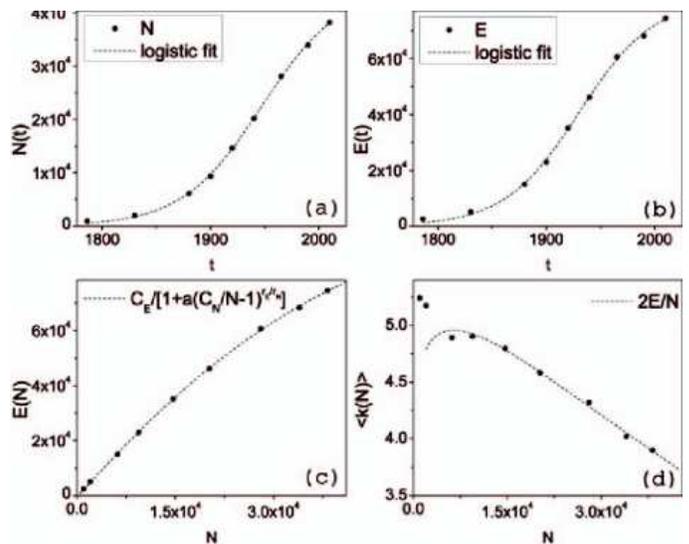}
 \caption{\label{f2} Dual network analysis. (a) Number of vertices $N(t)$ as a function of time. (b) Number of links $E(t)$ as a function of time. (c) Number of links $E(N)$ as a function of the number of vertices. (d) Average degree $<k(N)>$ as a function of the number of vertices.}
 \end{figure}  

\section{Time Evolution}
Using the HICN method defined above, we extract nine topological networks representing the evolution of the information space associated with London's street network over 224 years.  
In the top panels of Fig.\ref{f2}, we show the growth of the network as a function of time, with the number of vertices $N(t)$ plotted in the left panel and the number of links $E(t)$  in the right. 
Interestingly enough, we find that both functions can be fitted by a logistic function with a striking level of precision ($adjR^2>0.99$
\footnote{Calculating the goodness of fit for a non linear function is not a trivial issue, especially when the data is extracted by a dataset of the kind we use in this study, i.e. there are just a few points and no error bars \cite{error}. In order to quantify the goodness for the present fit we calculated the expected values for the fitted function and we considered the plot $f_{real}$ vs. $f_{estimated}$. Then we estimated the $adj.R^2$ derived by fitting the resulting plot with the function $y=x$.}
):
\begin{equation}\label{log}
f(t)=\frac{C}{1+e^{-r(t-t_0)}},
\end{equation} 
 where $r$  is the growth rate and $C$  the carrying capacity, while $t_0$  is the inflection point, that is $\partial^2 f /\partial t^2|_{t=t_0}=0$. 
 In \cite{masucci2} we demonstrated that the very same observation holds for the primary graph as well.
  This means that the dual graph of the London street network can be framed in a model of growth with competition for space, i.e. as a space filling phenomena  in a capacitated limit \cite{logistic}, where the London's \textit{green belt}, adopted in the 1950s, acts as bias and a constraint on  the free growth of the network.
 
Eq.\ref{log} allows us to forecast the growth of the network $E(N)$:
\begin{equation}\label{EN}
E(N)=\frac{C_E}{\left[1+a\left(\frac{C_N}{N}-1\right)^\frac{r_E}{r_N}\right]},
\end{equation}
where     $a=\exp[r_E(t_{0E}-t_{0N})]$ is constant, and the evolution of the average degree is $<k(N)>=2E/N$. 
Note that $C_E$ and $C_N$ are the capacities of the edges and nodes respectively while $r_E$ and $r_N$ are the respective rates of change for edges and nodes.
 These functions are plotted in the bottom panels of Fig.\ref{f2}. From the behaviour of the average degree, it is possible to trace how London's street network evolves from a more clustered topology in the information space, to a more tree-like one. 
 Assuming that the parameters of the logistic functions are stable, we can predict the asymptotic value of such quantities, i.e. $<k^\infty>=2E^\infty/N^\infty=2C_E/C_N\approx 3.62$.

\begin{figure}[!ht]\center
                \includegraphics[width=0.5\textwidth]{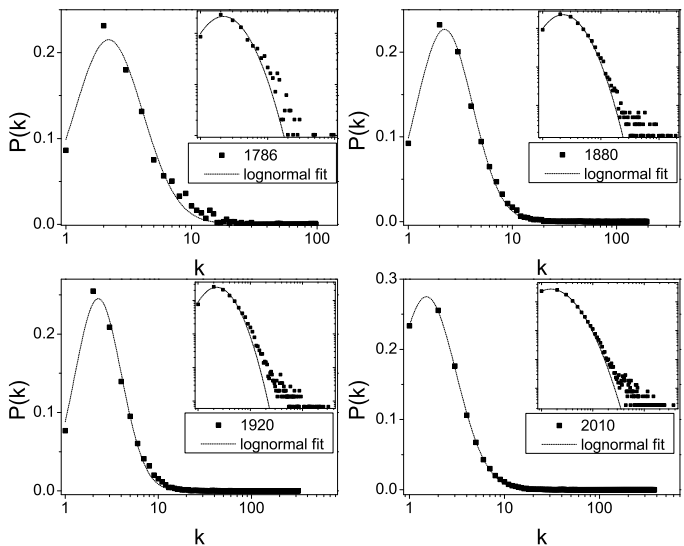}
 \caption{\label{f4}Degree distribution for the dual London street network for different time slices. In the insets the same plots are displayed in order to highlight the fat tail behaviour. }
 \end{figure}
 
\section{Topological Properties}

The topological properties of the network are well described by the connectivity distribution $P(k)$, which tells us the frequency of vertices having a certain number of neighbours, or connectivity, $k$. 
This measure has been applied in the analysis of cities in several studies  \cite{rosvaldual,portadual,masuccilond} and it has been concluded that it has a broad distribution and can be considered scale-free. As we mentioned before, we find that the scale-free behaviour is an artefact of the ICN principle.

In  Fig.\ref{f4}, we show the degree distribution of the network at different time slices. The plots are all well fitted by log-normal distributions ( $adjR^2>0.98$ ).
This observation is quite relevant, since the hypothesis that the dual space of road networks is scale free implies a self-organising dynamics in the information space \cite{dorog,newm}.
 Yet, the log-normal distributions we observe could be more accurately described as a multiplicative process  \textit{a la} Gibrat  \cite{gibrat}.

\begin{figure}[!ht]\center
                \includegraphics[width=0.5\textwidth]{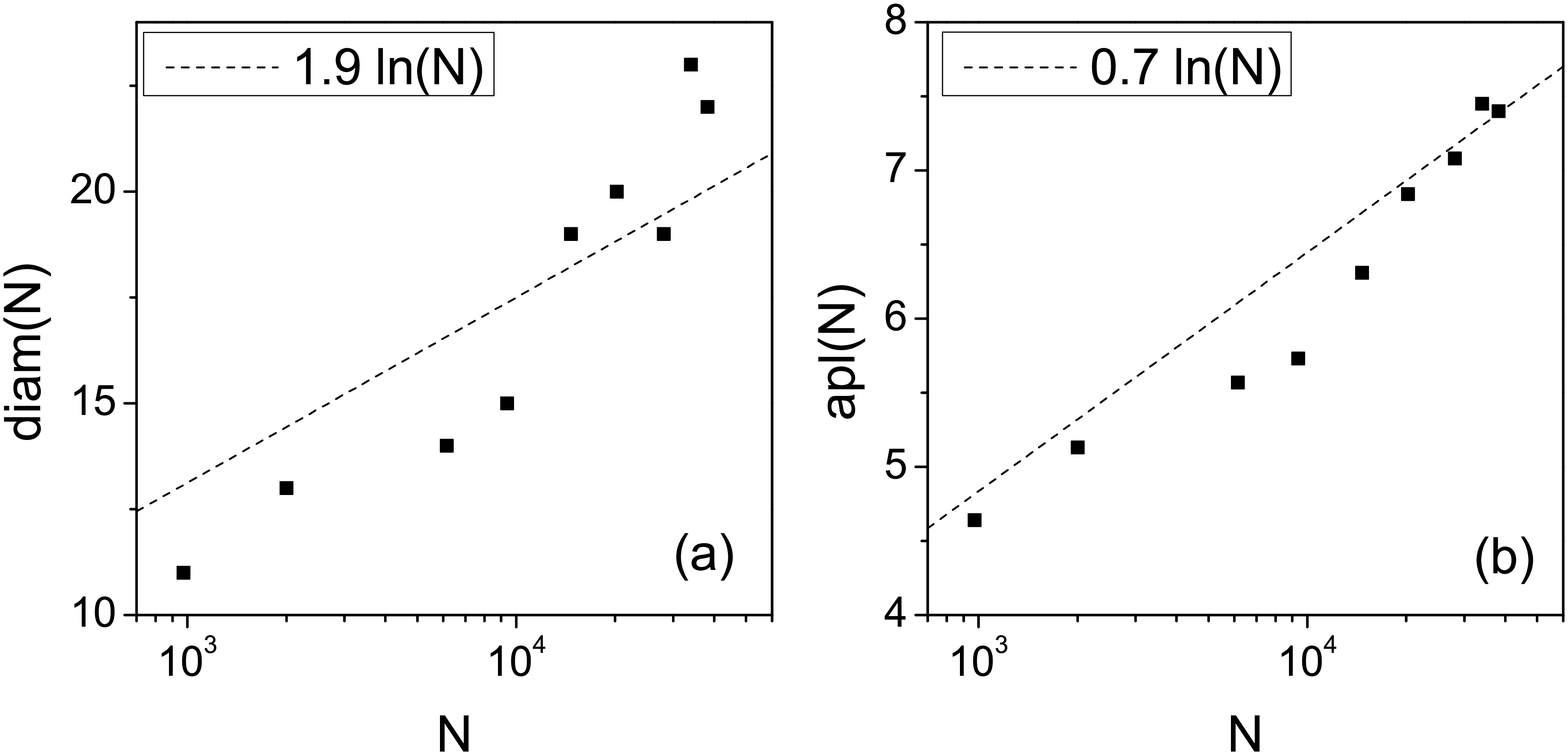}
 \caption{\label{f4b} (a) The diameter of the network $diam(N)$ as a function of the number of the vertices. (b) The average path length of the network $apl(N)$ as a function of the number of the vertices. }
 \end{figure}

In the left panel of Fig.\ref{f4b}, we show the diameter of the network $diam(N)$ as a function of the number of vertices. It is interesting to note that even if the networks are not scale free, they maintain some of the properties characteristic of self-organised networks. For instance, we see that the diameter of the network grows with the scale of the logarithm of its size,  a property typical of \textit{small world} networks \cite{watts}. This means that the city, despite its complexity, is easy to navigate. 
We speculate that the navigability of the city, as expressed by the logarithmic dependence of the diameter of the information space in respect to its size, is a constitutive property in the evolution of London's street networks.

\begin{figure}[!ht]\center
                \includegraphics[width=0.49\textwidth]{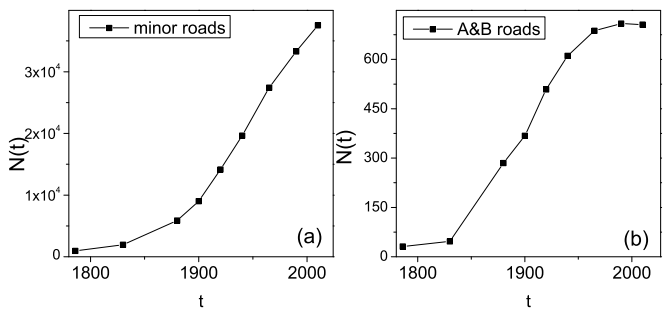}
 \caption{\label{f6} Left-panel: number of vertices $N(t)$ as a function of time for minor roads. Right-panel: number of vertices $N(t)$ as a function of time for class A and class B roads. }
 \end{figure}

In \cite{masucci2} we argued that the growth of a city is strongly hierarchical in the sense that the major roads mostly pre-date urbanisation, which is interpreted as a process of gradual filling of the blocks created by class A and class B roads with minor roads. 
This can be seen in Fig.\ref{f6}, where the number of roads is counted as a function of time for minor roads in the left panel and for class A and B roads in the right panel.
Even if class A and B roads absorb the main traffic flows in the city, their number is always at least two orders of magnitude less than the minor roads. 
It is also interesting to notice how the number of vertices for major roads at the last point - the 2010 network - is lower than the previous one - 1990 network. 
This dip in the line reflects the fact that the major roads undergo a number of complex phenomena related to splitting and, in this case, merging,  due to the completion of major road  construction executed in fragmented segments over time. 

\begin{figure}[!ht]\center
                \includegraphics[width=0.5\textwidth]{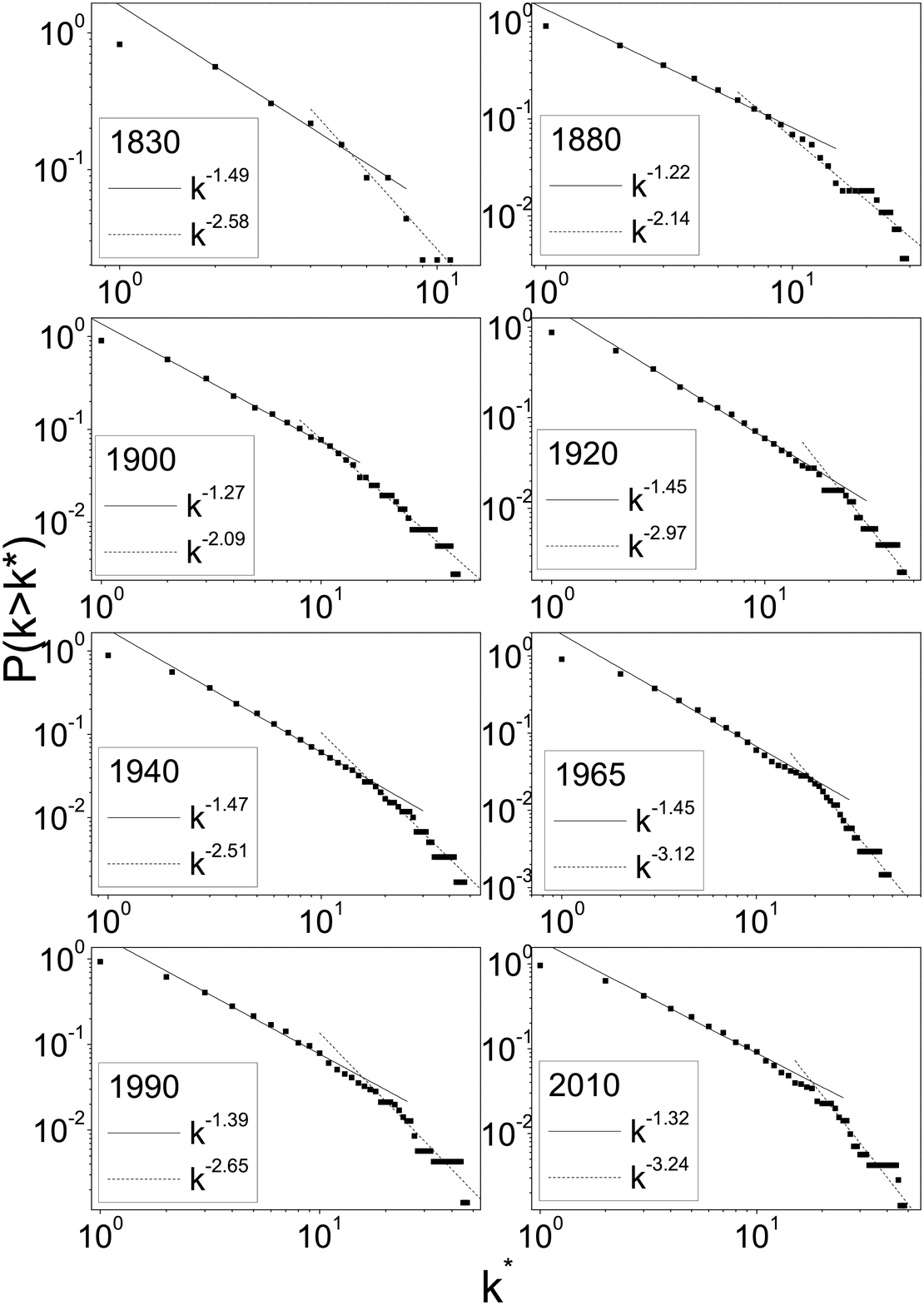}
 \caption{\label{f7} Cumulative degree distribution for the network formed by class A and B roads over time.  }
 \end{figure}

 In order to characterize the behaviour of  roads at a high level in the hierarchy, which form the backbone of the city, in Fig.\ref{f7} we show the cumulative degree distribution for the networks formed just for the major roads (A and B roads plus motorways). Interestingly enough, these distributions are not log-normal, but appear to follow a truncated power law or a double-Pareto distribution with a cut-off point around   $10<k<20$. This same behaviour was noticed already  in a study of the national road networks in the US, England and Denmark \cite{kalapala}. 
In particular, the exponent found in the upper tail of the distribution in England  in \cite{kalapala} is  comparable to those we find for London. This finding suggests that a truncated skew  distribution is a consistent characteristic  for major roads in the dual space. 

\begin{figure}[!ht]\center
                \includegraphics[width=0.5\textwidth]{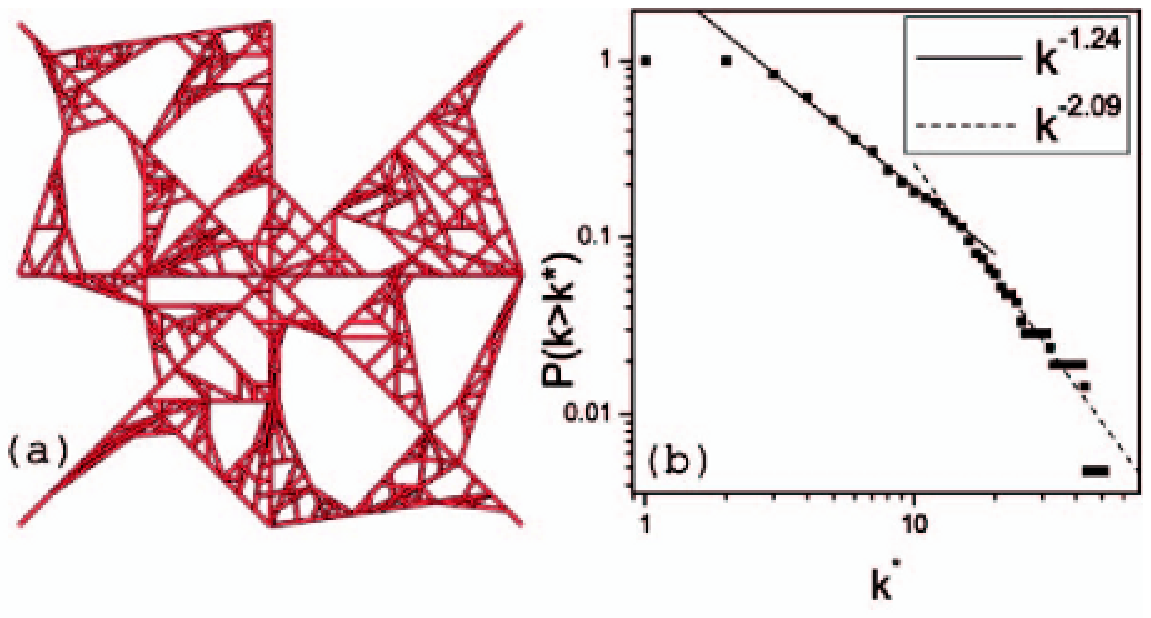}
                \includegraphics[width=0.5\textwidth]{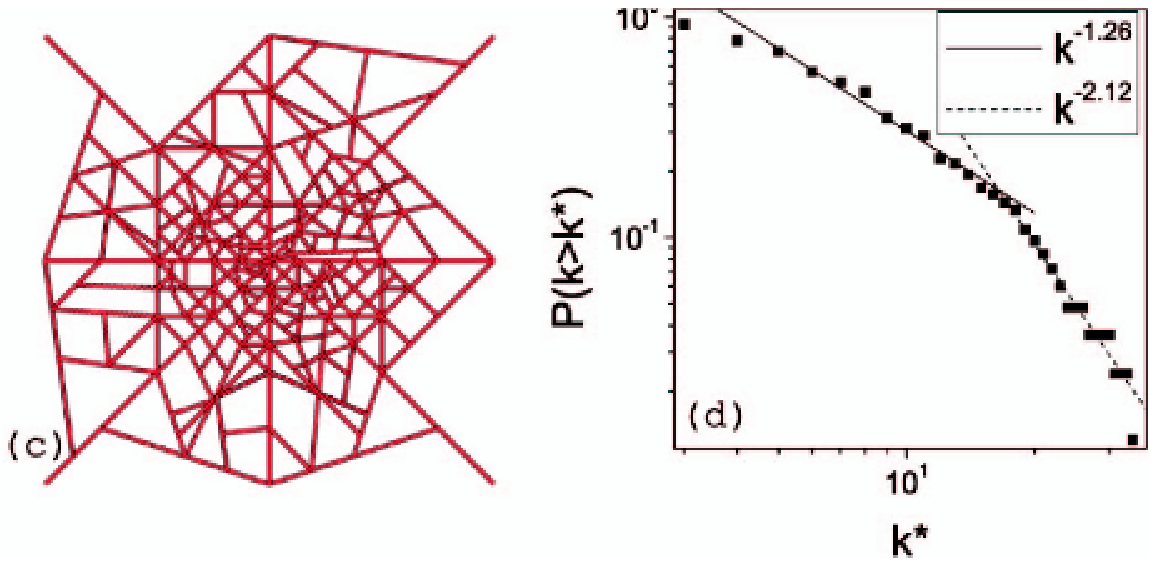}
 \caption{\label{f8} (Colours on-line) Top panels: Model1. Left panel: a realization of the model with 800 time-steps. Right panel: cumulative degree distribution measured for the model.
 Bottom panels: Model2. Left panel: a realization of the model with 450 time-steps. Right panel: cumulative degree distribution measured for the model.}
 \end{figure}

\section{A Space Filling Model}

Since city growth is a very complex phenomena, it has been mostly represented using algorithmic approaches \cite{mikefract,bartprl, slime2012}. These approaches consider street segment portions \cite{bartprl}, or at least street segments themselves \cite{masuccilond}, to be the basic elements in the evolution of street networks. However, our analysis questions the validity of this assumption. Street segments often do not emerge as constitutive elements of the network, but rather derive from the fragmentation of larger segments by their intersection with new roads. On the other hand, major roads, interpreted as vertices in the dual representation, often do not emerge as single roads, but are a result of merging of different roads. 

Here we  introduce a space filling model to explain the emergence of the double Pareto  distribution for the major roads (see Fig.\ref{f7}) \cite{reeds}. 
We generate roads in the dual space as a result of a merging process and the street segments in the metrical space as a result of fragmentation. The main idea of the model is that major roads serve to feed the city centre with resources (energy, materials, workers). Nodes (transport stations) emerge along these large roads, and new roads are created in order to connect these nodes to  existing intersections nearby. 

In Model 1, we start from a unit square area equally divided by 4 lines, all crossing in the centre of the square (see top-left panel of Fig.\ref{f8}). Each intersection splits a road in individual street segments. At each time step we randomly pick up a street segment $i$, of length $l_i>0.05$, with probability $P_i$ proportional to its length $l_i$, $P_i= l_i/\sum_jl_j$. Then we create a new node at the middle point of this street segment and we connect this new vertex to the closest visible intersection, thus adding  a new street segment. The model stops when there are no segments of length larger than 0.05 anymore, i.e. when the available space is filled.  This simple model, even if not particularly realistic, generates in the dual space a similar distribution to that found in London (see right panel of Fig.\ref{f8}). Other ingredients can be added to reproduce shapes that are more similar to the structures we measure in reality, such as a maximum degree for the intersections, a bias on the shape factor, and so on.

In the bottom panels of the same figure we show a more realistic realization of the model, based on the same space filling principles.
In Model 2, we start with the same initial conditions as in the previous model and we proceed in the same way.  At each time step we randomly pick up a street segment, with probability proportional to its length. Then we create a new node in the middle point of this street segment and we connect it to the closest visible intersection, adding a new street segment. The difference is that this time we put a bias on the areas of the generated polygons, and on the angles at the intersections. In particular we assume that the area of the polygons formed by the street segments $A(r)$ is an increasing function of the distance from the centre $r$, i.e. $A(r)>0.05e^{-\frac{1}{r}}$, and that the angles at the intersections are greater than $\pi/4$. Model 2 gives a more realistic morphology to the resulting street network and at the same time preserves a double-Pareto degree distribution with exponents similar to those found in reality.

\section{Conclusions}
 
A growing number of studies have demonstrated that the dual representation of a street network is fundamental to understand the dynamics of transportation networks in cities. Here we show some of the limitations of the existing methods for extracting such a network from planar graphs and we propose a new method, the HICN, that is based on geometrical and hierarchical features, which can be easily obtained from digital maps available in GIS   \cite{mikegis}. 

In the ICN method we use one parameter, the angular threshold, which we choose to be $\pi/2$. 
The choice of this value is mainly arbitrary and resides on the fact that in many real cases roads which intersect forming an acute angle are not considered as the same road in the information space. 
One could argue about the sensitivity of the resulting dual graph to this parameter. 
It is to say that in the hierarchical approach there are just a minority of roads that intersect with an angle minor than the one we choose. 
In fact, for the case of London, the cases where the street segments form angles smaller than $\pi/2$ belong generally to different road hierarchies. 
As an example, in the 2010 London's map the angular threshold of $\pi/2$ applies just in the 2\% of the street segments.
In the light of this, we can say that the choice of an angular threshold minor than $\pi/2$ would not affect the global topological results of the network.  
Choosing a larger angle as a threshold would fragment the roads, pushing this research toward the \textit{space syntax} approach, which considers two street segments belonging to the same road when they are on the same line of sight \cite{spsynt}. 
However, such an approach is beyond our line of research. 
In the larger limit, i.e. choosing the threshold to be $\pi$, the resulting dual network would be a so called line graph \cite{lineg}.

Using this methodology, we explore the growth of London's street network in the information space during a period of 224 years. The long time range covered by the dataset allows us to look for stable statistical properties of the network over time. We reach some unexpected conclusions regarding the logistic laws governing the growth of the city and the analytical prediction of the growth of the network and its average degree. Further, we find that the topology of the information space can be described by log-normal distributions and not by scale-free distributions as it has been previously argued. This observation reframes the interpretation of the information space in a dynamics following Gibrat. 

Moreover, we show that the topology of the major roads for London is comparable with the results of previous studies of large scale road networks, i.e. it represents a truncated skew distribution which resembles a double-Pareto distribution. Unfortunately, since such distribution appears for less than two decades, it is difficult to assess whether it is a power law or not. However we show that this kind of behaviour can be  reproduced by a space filling model. A novelty of this model with respect to others is that the final planar graph is the result of the fragmentation processes acting on longer segments and that roads can be generated by a process of merging street segments created in different times. This approach produces more realistic results than the one that considers the resulting planar graph representing a city as a mere addition of street segments to one or more initial seeds.

The results shown here are based solely on the analysis of the evolution of Greater London's road network. Our previous analysis  \cite{masucci2} indicated that the geometrical properties of a city strongly depend on the adopted spatial development policies. It is therefore necessary to assess the validity of the conclusions propounded in this paper in relation to other urban systems, in order to find out if universal properties for the evolution of urban systems can be established. 

\bibliographystyle{apsrev4-1}
\bibliography{dual}

 \end{document}